\RequirePackage{lineno}  
\documentclass[aps,prl,showpacs,showkeys,twocolumn,superscriptaddress]{revtex4} % UTOBS

\usepackage{graphicx}
\newcommand{\ET}{{\rm {\it E}}_{T}}
\newcommand{\MET}{\mbox{$\raisebox{.3ex}{$\not$}\ET$}}
\newcommand{\lsim}{~\rlap{$<$}{\lower 1.0ex\hbox{$\sim$}}}

\begin{document}

% LINENO CTOBS
%\pagewiselinenumbers  


%\preprint{CDF/PUB/EXOTIC/PUBLIC/9080} %Comment this out before shipping out

%\preprint{Version 1.5}          %Comment this out before shipping out

%%% AH \title{ Search for charged Higgs boson in \ttbar decay  products}
\title{Search for the Higgs boson produced in association with $Z \to \ell^+\ell^-$ in
$p\bar{p}$ collisions at $\sqrt{s}=$ 1.96 TeV}

%\input{July_2007_Authors_Visitors}

%This gets replaced by authors and institutions
%\vspace{0.5in}

%\input{Oct_2006_Authors_alt}
\affiliation{Institute of Physics, Academia Sinica, Taipei, Taiwan 11529, Republic of China} 
\affiliation{Argonne National Laboratory, Argonne, Illinois 60439} 
\affiliation{University of Athens, 157 71 Athens, Greece} 
\affiliation{Institut de Fisica d'Altes Energies, Universitat Autonoma de Barcelona, E-08193, Bellaterra (Barcelona), Spain} 
\affiliation{Baylor University, Waco, Texas  76798} 
\affiliation{Istituto Nazionale di Fisica Nucleare Bologna, $^t$University of Bologna, I-40127 Bologna, Italy} 
\affiliation{Brandeis University, Waltham, Massachusetts 02254} 
\affiliation{University of California, Davis, Davis, California  95616} 
\affiliation{University of California, Los Angeles, Los Angeles, California  90024} 
\affiliation{University of California, San Diego, La Jolla, California  92093} 
\affiliation{University of California, Santa Barbara, Santa Barbara, California 93106} 
\affiliation{Instituto de Fisica de Cantabria, CSIC-University of Cantabria, 39005 Santander, Spain} 
\affiliation{Carnegie Mellon University, Pittsburgh, PA  15213} 
\affiliation{Enrico Fermi Institute, University of Chicago, Chicago, Illinois 60637} 
\affiliation{Comenius University, 842 48 Bratislava, Slovakia; Institute of Experimental Physics, 040 01 Kosice, Slovakia} 
\affiliation{Joint Institute for Nuclear Research, RU-141980 Dubna, Russia} 
\affiliation{Duke University, Durham, North Carolina  27708} 
\affiliation{Fermi National Accelerator Laboratory, Batavia, Illinois 60510} 
\affiliation{University of Florida, Gainesville, Florida  32611} 
\affiliation{Laboratori Nazionali di Frascati, Istituto Nazionale di Fisica Nucleare, I-00044 Frascati, Italy} 
\affiliation{University of Geneva, CH-1211 Geneva 4, Switzerland} 
\affiliation{Glasgow University, Glasgow G12 8QQ, United Kingdom} 
\affiliation{Harvard University, Cambridge, Massachusetts 02138} 
\affiliation{Division of High Energy Physics, Department of Physics, University of Helsinki and Helsinki Institute of Physics, FIN-00014, Helsinki, Finland} 
\affiliation{University of Illinois, Urbana, Illinois 61801} 
\affiliation{The Johns Hopkins University, Baltimore, Maryland 21218} 
\affiliation{Institut f\"{u}r Experimentelle Kernphysik, Universit\"{a}t Karlsruhe, 76128 Karlsruhe, Germany} 
\affiliation{Center for High Energy Physics: Kyungpook National University, Daegu 702-701, Korea; Seoul National University, Seoul 151-742, Korea; Sungkyunkwan University, Suwon 440-746, Korea; Korea Institute of Science and Technology Information, Daejeon, 305-806, Korea; Chonnam National University, Gwangju, 500-757, Korea} 
\affiliation{Ernest Orlando Lawrence Berkeley National Laboratory, Berkeley, California 94720} 
\affiliation{University of Liverpool, Liverpool L69 7ZE, United Kingdom} 
\affiliation{University College London, London WC1E 6BT, United Kingdom} 
\affiliation{Centro de Investigaciones Energeticas Medioambientales y Tecnologicas, E-28040 Madrid, Spain} 
\affiliation{Massachusetts Institute of Technology, Cambridge, Massachusetts  02139} 
\affiliation{Institute of Particle Physics: McGill University, Montr\'{e}al, Canada H3A~2T8; and University of Toronto, Toronto, Canada M5S~1A7} 
\affiliation{University of Michigan, Ann Arbor, Michigan 48109} 
\affiliation{Michigan State University, East Lansing, Michigan  48824}
\affiliation{Institution for Theoretical and Experimental Physics, ITEP, Moscow 117259, Russia} 
\affiliation{University of New Mexico, Albuquerque, New Mexico 87131} 
\affiliation{Northwestern University, Evanston, Illinois  60208} 
\affiliation{The Ohio State University, Columbus, Ohio  43210} 
\affiliation{Okayama University, Okayama 700-8530, Japan} 
\affiliation{Osaka City University, Osaka 588, Japan} 
\affiliation{University of Oxford, Oxford OX1 3RH, United Kingdom} 
\affiliation{Istituto Nazionale di Fisica Nucleare, Sezione di Padova-Trento, $^u$University of Padova, I-35131 Padova, Italy} 
\affiliation{LPNHE, Universite Pierre et Marie Curie/IN2P3-CNRS, UMR7585, Paris, F-75252 France} 
\affiliation{University of Pennsylvania, Philadelphia, Pennsylvania 19104} 
\affiliation{Istituto Nazionale di Fisica Nucleare Pisa, $^q$University of Pisa, $^r$University of Siena and $^s$Scuola Normale Superiore, I-56127 Pisa, Italy} 
\affiliation{University of Pittsburgh, Pittsburgh, Pennsylvania 15260} 
\affiliation{Purdue University, West Lafayette, Indiana 47907} 
\affiliation{University of Rochester, Rochester, New York 14627} 
\affiliation{The Rockefeller University, New York, New York 10021} 

\affiliation{Istituto Nazionale di Fisica Nucleare, Sezione di Roma 1, $^v$Sapienza Universit\`{a} di Roma, I-00185 Roma, Italy} 

\affiliation{Rutgers University, Piscataway, New Jersey 08855} 
\affiliation{Texas A\&M University, College Station, Texas 77843} 
\affiliation{Istituto Nazionale di Fisica Nucleare Trieste/\ Udine,
$^w$University of Trieste/\ Udine, I-33100 Udine, I-34100 Trieste, Italy} 
\affiliation{University of Tsukuba, Tsukuba, Ibaraki 305, Japan} 
\affiliation{Tufts University, Medford, Massachusetts 02155} 
\affiliation{Waseda University, Tokyo 169, Japan} 
\affiliation{Wayne State University, Detroit, Michigan  48201} 
\affiliation{University of Wisconsin, Madison, Wisconsin 53706} 
\affiliation{Yale University, New Haven, Connecticut 06520} 
\author{T.~Aaltonen}
\affiliation{Division of High Energy Physics, Department of Physics, University of Helsinki and Helsinki Institute of Physics, FIN-00014, Helsinki, Finland}
\author{J.~Adelman}
\affiliation{Enrico Fermi Institute, University of Chicago, Chicago, Illinois 60637}
\author{T.~Akimoto}
\affiliation{University of Tsukuba, Tsukuba, Ibaraki 305, Japan}
\author{M.G.~Albrow}
\affiliation{Fermi National Accelerator Laboratory, Batavia, Illinois 60510}
\author{B.~\'{A}lvarez~Gonz\'{a}lez}
\affiliation{Instituto de Fisica de Cantabria, CSIC-University of Cantabria, 39005 Santander, Spain}
\author{S.~Amerio$^u$}
\affiliation{Istituto Nazionale di Fisica Nucleare, Sezione di Padova-Trento, $^u$University of Padova, I-35131 Padova, Italy} 

\author{D.~Amidei}
\affiliation{University of Michigan, Ann Arbor, Michigan 48109}
\author{A.~Anastassov}
\affiliation{Northwestern University, Evanston, Illinois  60208}
\author{A.~Annovi}
\affiliation{Laboratori Nazionali di Frascati, Istituto Nazionale di Fisica Nucleare, I-00044 Frascati, Italy}
\author{J.~Antos}
\affiliation{Comenius University, 842 48 Bratislava, Slovakia; Institute of Experimental Physics, 040 01 Kosice, Slovakia}
% ...

\collaboration{CDF Collaboration\footnote{With visitors from $^a$Universiteit Antwerpen, B-2610 Antwerp, Belgium, 
$^b$Chinese Academy of Sciences, Beijing 100864, China, 
$^c$University of Bristol, Bristol BS8 1TL, United Kingdom, 
$^d$University of California Irvine, Irvine, CA  92697, 
$^e$University of California Santa Cruz, Santa Cruz, CA  95064, 
$^f$Cornell University, Ithaca, NY  14853, 
$^g$University of Cyprus, Nicosia CY-1678, Cyprus, 
$^h$University College Dublin, Dublin 4, Ireland, 
$^i$University of Edinburgh, Edinburgh EH9 3JZ, United Kingdom, 
$^j$Universidad Iberoamericana, Mexico D.F., Mexico, 
$^k$University of Manchester, Manchester M13 9PL, England, 
$^l$Nagasaki Institute of Applied Science, Nagasaki, Japan, 
$^m$University de Oviedo, E-33007 Oviedo, Spain, 
$^n$Queen Mary, University of London, London, E1 4NS, England, 
$^o$Texas Tech University, Lubbock, TX  79409, 
$^p$IFIC(CSIC-Universitat de Valencia), 46071 Valencia, Spain,
$^x$Royal Society of Edinburgh/Scottish Executive Support Research Fellow,
$^y$Istituto Nazionale di Fisica Nucleare, Sezione di Cagliari, 09042 Monserrato (Cagliari), Italy 
}}
\noaffiliation

\author{The CDF Collaboration} % This will get replaced with the CDF author list. See the GP webpage to get it.

\date{\today}

\begin{abstract}

 We present a search for the Higgs boson in the process $q\bar{q} \to ZH \to
 \ell^+\ell^- b\bar{b}$. The analysis uses an integrated luminosity of 1 fb$^{-1}$
 of $p\bar{p}$ collisions produced at $\sqrt{s} =$ 1.96 TeV and accumulated
 by the upgraded Collider Detector at Fermilab (CDF II).  We employ
 artificial neural networks both to correct jets mismeasured in the
 calorimeter, and to distinguish the signal kinematic distributions from those of the
 background. We see no evidence for Higgs boson production, and set 95\%
 C.L. upper limits on $\sigma_{ZH} \cdot {\cal B}(H \to b\bar{b}$),
 ranging from 1.5 pb to 1.2 pb for a Higgs boson mass ($m_H$) of 110 to 150
GeV/$c^2$. 
%The improvement in limits using these approaches is equivalent
%to an increase in luminosity by a factor of 3.3, compared with fitting the
%uncorrected dijet mass alone. 

\end{abstract}

%
% 13.38.Dg : Decays of Z bosons
% 13.85.Qk : Inclusive production with identified leptons, photons, or other nonhadronic particles
% 14.70.Hp : Z bosons
% 14.80.Bn : standard model Higgs bosons
% 14.80.Cp : non-standard-model Higgs bosons

\maketitle


%
%    INTRODUCTIONX
%

The Higgs boson is the only particle predicted by the standard model (SM) of
particle physics which has not yet been discovered. It is the physical
manifestation of the mechanism which provides mass to fundamental
particles~\cite{Higgs:1964ia,Guralnik:1964eu}.  Direct searches have excluded
the SM Higgs boson for masses $m_H <$ 114.4 GeV/$c^2$ at the 95\%
C.L.~\cite{Barate:2003sz}.  The Higgs boson mass is indirectly constrained
from precise electroweak measurements to $m_H$ = 76$^{+33}_{-24}$
GeV/$c^2$~\cite{Alcaraz:2007ri}.  A number of extensions to the SM predict a
SM-like Higgs boson, in particular Ref.~\cite{Roszkowski:2006mi} predicts a
SM-like Higgs boson with 68\% posterior probability to be between 115.4 and
120.4 GeV/$c^2$.  Only the Tevatron collider experiments are currently
capable of extending the limits on a Higgs boson for $m_H >$ 114.4 GeV/$c^2$.

This Letter presents the first CDF II search for a Higgs boson in the process 
$p\bar{p}\to Z^{*}\to ZH\to \ell^+\ell^- b\bar{b}$, where $\ell$ is $e$ or
$\mu$, with a dataset of 1 fb$^{-1}$, almost three times that of the previously
reported analysis~\cite{Abazov:2007pj}.  
%This is the first run II Higgs
%analysis which makes use of multivariate techniques, in this case two
%artificial neural networks (ANN), one which corrects jet energies, and one
%which discriminates the kinematic distributions of the Higgs signal from
%those of background. 
CDF and D\O \ have previously presented Higgs boson searches in other decay
modes~\cite{Aaltonen:2007wx,Aaltonen:2008mi,Abulencia:2006aj,Abazov:2005un,Abazov:2004jy,
Abazov:2006pt,Abazov:2006hn}. 

CDF II \cite{Acosta:2004yw} is a general purpose detector. Its
coordinate system and quantities used throughout this paper are defined in
Ref.~\cite{coord}. At its center is a cylindrical silicon detector
which tracks charged particles from a radius of 1.35 to 29 cm for $|\eta|
\lsim$ 2.  Around this is a cylindrical wire drift chamber which tracks
charged particles from 43 to 132 cm for $|\eta| \lsim$ 1.3.  A superconducting
solenoid surrounds the tracking volume providing a 1.4 T magnetic field for
momenta measurements.  Segmented electromagnetic and hadronic sampling
calorimeters surrounding the solenoid measure energies of interacting
particles with $|\eta| <$ 3.6.  A system of drift chambers and scintillation
counters outside the calorimeters detect muon candidates for $|\eta| <$ 1.5.
%Gas Cherenkov counters at the forward
%regions of 3.7 $< |\eta| <$ 4.7 measure the inelastic $p\bar{p}$ collision
%rate per bunch crossing to determine the beam luminosity.

At the Tevatron the cross section for $q\bar{q} \to Z^{*} \to ZH$ production for a Higgs
boson with mass $m_H =$ 115 GeV/$c^2$ is 1.04 pb~\cite{Catani:2003zt}, and
the branching ratio ${\cal B}(H\to b\bar{b})$ is
73\%~\cite{Djouadi:1997yw}. To identify candidate $ZH$ events, we first
search for $Z$ candidates decaying to electron or muon pairs.  The full
selection criteria are described in Ref.~\cite{Efron:2007zz}; the most
salient features are described here.  
%At least one lepton must pass trigger 
%requirements.  
%The inclusive electron and muon triggers collect events with
%95 $\pm$ 1 \% efficiency 
Events are collected using a trigger
which identifies a primary electron (muon) with $E_T >$ 18 GeV ($p_T
>$ 18 GeV/$c$) within the central region $|\eta| <$ 1.0.  The requirements for
the second electron are relaxed to $E_T >$ 10 GeV in the central region, and
maintained at 18 GeV for electrons with 1.0 $< |\eta| <$ 2.4. The second muon
must have $p_T >$ 10 GeV/$c$.  Energy deposits from leptons must be isolated
from other energy deposits within $\Delta R <$ 0.4.
%We develop a lepton selection which is
%looser than previously reported high $p_T$ lepton measurements
%\cite{top,higgs} in order to maintain as many real $Z$ events as possible. 
%The trigger electron requirements are $E_T
%>$ 18 GeV, a track with $p_T >9$~GeV/$c$ matched to its calorimeter cluster,
%and ``isolated'' such that additional calorimeter energy deposited within
%$\Delta R <$ 0.4 around the lepton is less than 10\% of the leptons
%transverse energy. For the trigger muon, the track must have $p_T >$ 20
%GeV/$c$ and well-matched signals in the muon chambers. It must be isolated,
%and have a low energy deposit in the hadronic calorimeter consistent with the
%expectation for a muon.  The second electron must be isolated. For electrons
%with $|\eta| <$ 1, $E_T$ and $p_T$ are relaxed to 10 GeV and 5 GeV/$c$.
%Electrons with 1.0 $< |\eta| <$ 2.4 are also accepted but with no track
%requirement due to the decreased tracking efficiency in this region.  The
%loose muon must be isolated, have $p_T >10$~GeV/$c$, and have an expected
%signal in the hadronic calorimeter, but it is not required to have a signal
%in the muon chambers. 

From the measured lepton energies and momenta, we reconstruct the invariant
mass of the $Z$ candidate and require it to be between 76 and 106
GeV/$c^2$. This requirement is 92\% efficient for real $ZH$ candidates, but
helps remove non-$Z$ backgrounds.  We require oppositely charged leptons for
muon pairs and for electron pairs when the second electron has $|\eta|<$1.0
due to improved tracking efficiency in the central region.
%(BH: really only 90\%??? Is this for Z's or for Z's in ZH?) 

Higgs boson candidates are then selected by requiring a jet with $E_T >$ 25 GeV and
an additional jet with $E_T >$ 15 GeV, both with $|\eta| <$ 2.0.  Jets are
corrected for calorimeter response, multiple interactions, and energy loss in
the uninstrumented detector regions~\cite{Bhatti:2005ai}.  To enhance signal
significance, we implement an algorithm to identify the decay of a long-lived
hadron containing a $b$ quark by reconstructing a significantly displaced
secondary vertex ~\cite{Acosta:2004hw,Neu:2006rs}. The efficiency is 40 to
50\% for $b$-quark jets and 1 to 3\% for $u$, $d$, $s$, or $g$ (light
parton, l.p.) jets.  We consider events in which one or both jets are ``tagged''
by this algorithm.

Backgrounds originating from $W$, $Z$, and $t\bar{t}$ production are
determined using leading order Monte Carlo (MC) calculations, normalized to
next to leading order, followed by a detailed simulation of the CDF II
detector. We model the $Z +b\bar{b}$, $Z +c\bar{c}$ and $Z$ + l.p. processes
by first producing the exact leading order multiparton final states with the
{\sc alpgen}~\cite{Mangano:2002ea} MC program, and then using the {\sc
herwig}~\cite{Corcella:2002jc} MC program to model the hadronization and parton
showering. In addition we use an inclusive {\sc
pythia}~\cite{Sjostrand:2001yu} $Z$ MC sample to compare with the observed
data and evaluate systematic uncertainties.  We model $ZZ$, $ZW$, and
$t\bar{t}$ background contributions using the {\sc pythia} MC program.  A
``fake lepton'' background arises from jets being misidentified as leptons,
and we estimate this contribution from observed data~\cite{Abulencia:2006ce}.
The contribution from false tags of l.p. jets is evaluated from data by
applying a parametrization of the false tagging rate to jets passing $E_T$
and $\eta$ requirements~\cite{Acosta:2004hw,Neu:2006rs}.  The acceptance,with
statistical uncertainties, of $ZH\to
\ell^+\ell^- b\bar{b}$ events is (10.8 $\pm$ 0.1)\% for $m_H = 120$ GeV/$c^2$
and is evaluated using the {\sc pythia} MC program followed by a detailed CDF
II simulation for Higgs boson masses from 110 to 150 GeV/$c^2$.

\begin{table}[h]
\caption{Expected and observed numbers of events in 1 fb$^{-1}$ for electron and muon
decay modes combined, compared to the expected $ZH$ signal for $m_H =$ 120
GeV/$c^2$.}
\label{tab:bkgsig}
\begin{center}
\begin{tabular}{lccc}            
\hline
\hline
  Sample  &      Single-tagged & Double-tagged \\
\hline
% This table belongs to Jonathan :)
$Z +b\bar{b}$    & $35.1 \pm 14.6$  &    $6.3 \pm 2.5$  \\
$Z +c\bar{c}$    & $21.8 \pm 8.5$  &    $1.0 \pm 0.4$  \\
$Z +$l.p.    & $32.3 \pm 5.5$  &    $1.0 \pm 0.2$  \\
   $t\bar{t}$    & $5.2 \pm 1.0$  &    $2.8 \pm 0.6$   \\
       $ZZ$      & $4.0 \pm 0.8$  &    $1.3 \pm 0.3$  \\
       $ZW$      & $1.2 \pm 0.2$  &    $0.04 \pm 0.01$  \\
  Non-$Z$        & $1.9 \pm 1.4$  &    $0.2 \pm 0.2$  \\
\hline                         
Expected         & $101.5 \pm 32.0$  &    $12.7 \pm 4.1$  \\ 
\hline
Observed         & $100$           &    $11$   \\ \hline
\hline
   $ZH$    & $0.44$  &    $0.23$  \\
\hline
\hline
\end{tabular}
\end{center}
\end{table}
Table \ref{tab:bkgsig} shows the expected background and signal contributions
with systematic uncertainties in 1 fb$^{-1}$ of data compared to the number
observed after dividing events into those with only one $b$-tag
(single-tagged), and those with exactly two $b$-tags
(double-tagged). Electron decay modes account for 60\% of the total expected
and observed events.

%We take several steps to improve discrimination of the $ZH$ signal from the
%backgrounds. 
The $Z$+jets system is only expected to have $\MET$ due to jet
mismeasurement, either from the limited calorimeter resolution,
uninstrumented regions, or from the semi-leptonic decay of the jets.
Therefore, we correct the jet energies using a multilayer perceptron
Artificial Neural Network (ANN)~\cite{DBLP:journals/nn/HornikSW89} which
uses the missing transverse energy $\MET$ vector projected onto those of the
jets in order to determine individual scale factors for each jet.  The result
is a dijet mass resolution improvement from 18\% to 11\% ~\cite{Efron:2007zz}
which improves $ZH$ signal discrimination from the backgrounds.

% while
%lowering the amount of background expected around the Higgs dijet resonance
%peak as in Ref.~\cite{Efron:2007zz}.  

%The dijet mass distribution for events
%with one $b$ tag is shown in Fig.~\ref{fig:dijet}.

%We find agreement between data and MC
%for this correction function by testing it with 3000 events passing all
%selection criteria except the $b$-tag requirements, and using the $Z$ $p_T$
%as a probe to measure the energy resolution of the dijet system. 

% We determine a subset of these variables by
%sequentially adding in single variables and retraining an ANN to obtain the
%smallest average error for correctly classifying the events as either $ZH$,
%$Z$+jets, or $t\bar{t}$.  We stop adding in variables once there is no
%variable which can improve the training error by 1\%. 

\begin{figure}[htbp]
\includegraphics[width=0.45\textwidth]{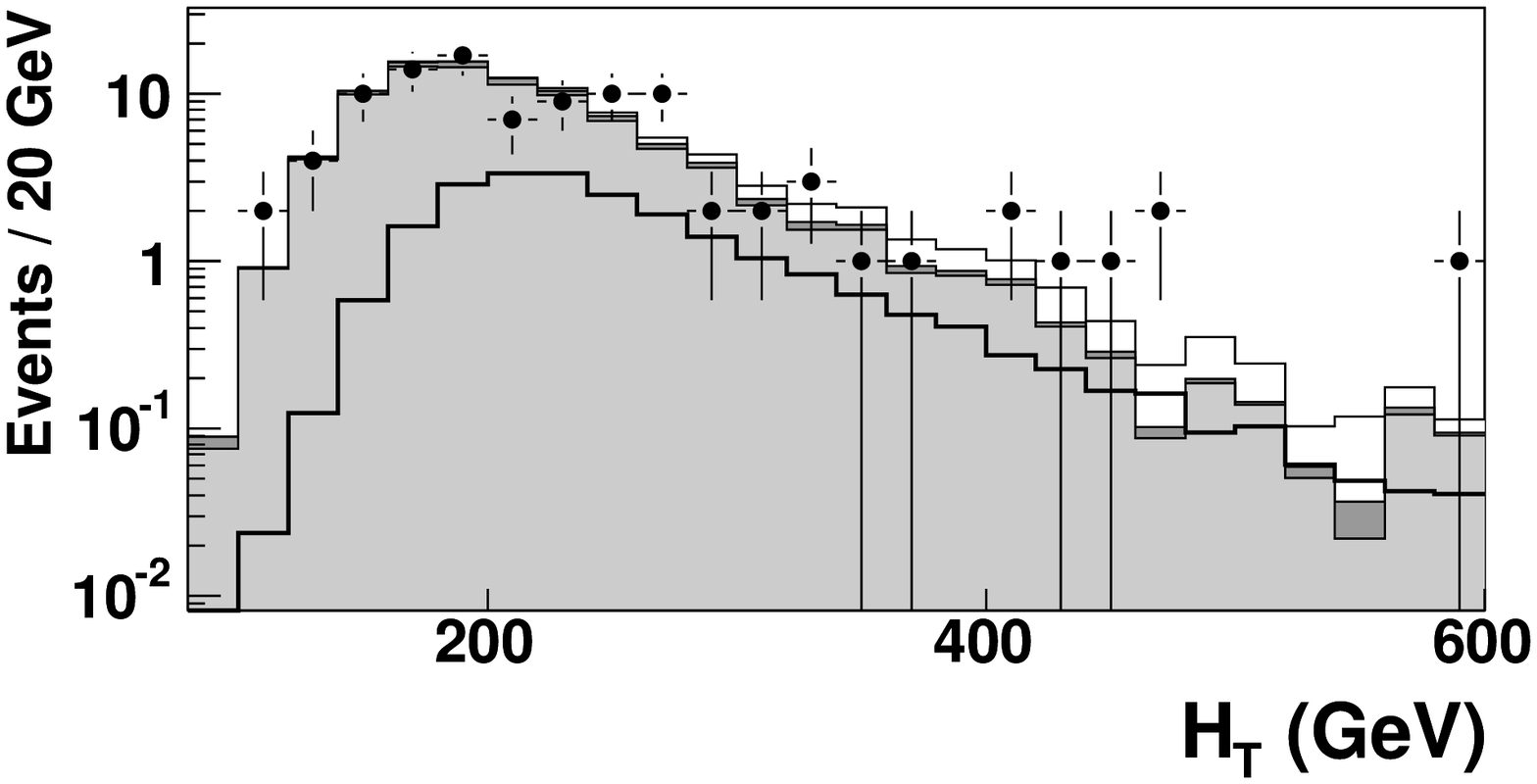}\\
\includegraphics[width=0.45\textwidth]{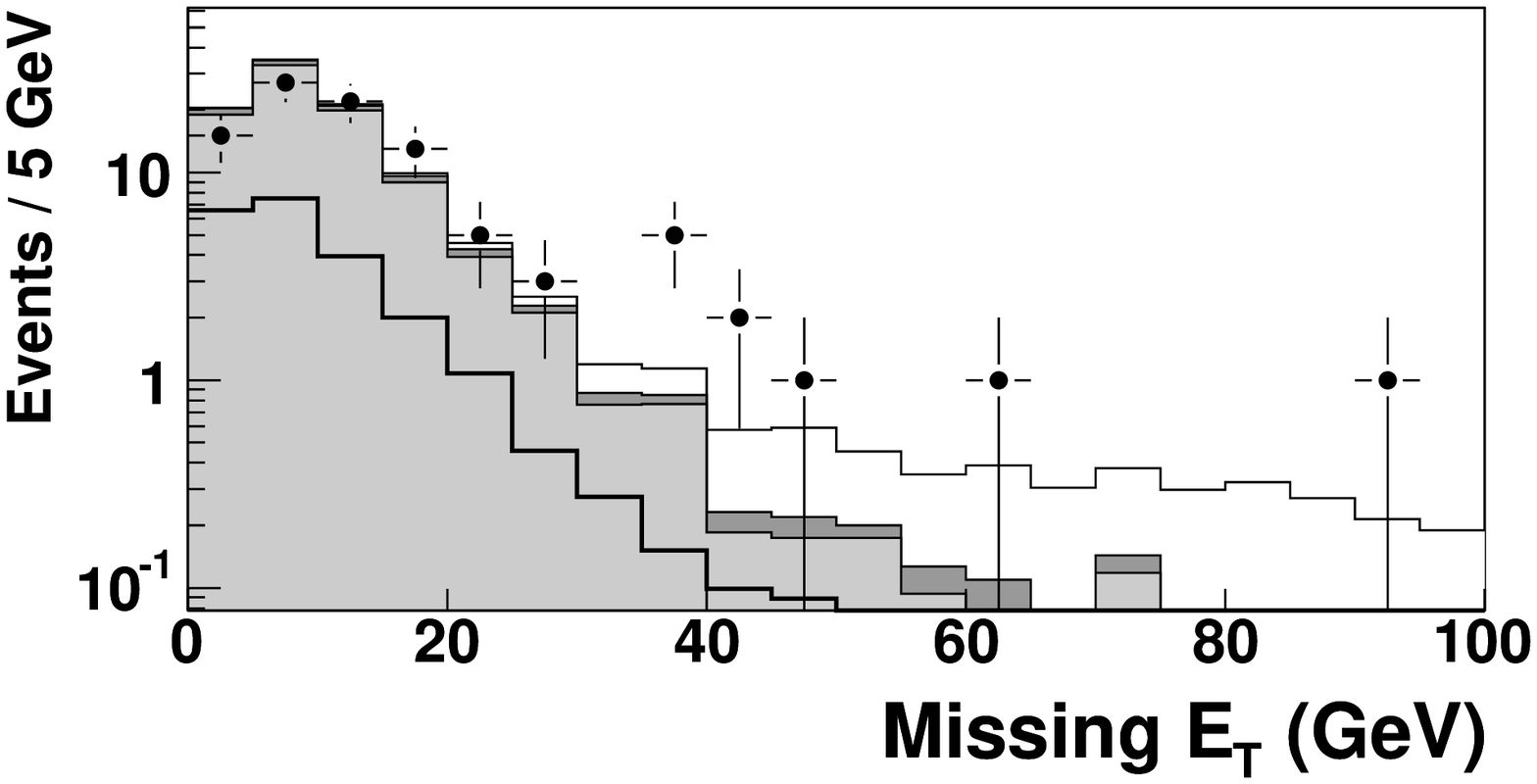} \\
\includegraphics[width=0.45\textwidth]{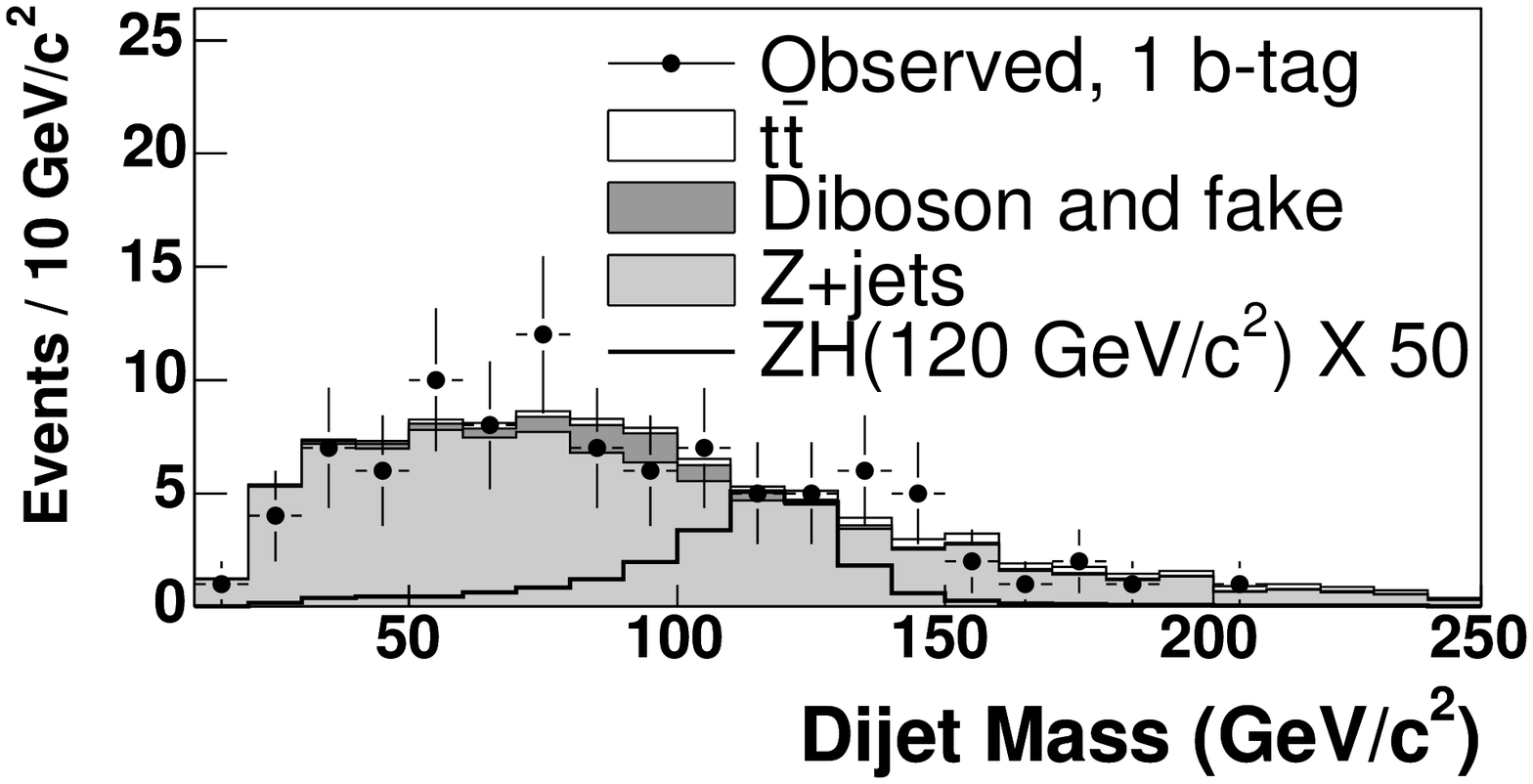}
\caption{\label{fig:dijet}
Expected and observed distributions for the three most important inputs to
the signal discriminating ANN shown after ANN jet corrections have been
applied, for events with 1 $b$-tag.  
%The top plot is scalar sum of the
%transverse energies of the two leptons and two jets ($H_T$), the central plot
%is the missing transverse energy, and the bottom plot is the dijet mass of
%the Higgs candidate.
}
\end{figure}

To achieve a greater separation of signal and background we employ an
additional ANN implemented with {\sc jetnet}~\cite{Peterson:1993nk} to
distinguish the kinematics of the signal from those of the backgrounds.  
%An analysis of the combination of ANNs instead of the dijet mass distribution
%provides a factor of 1.8 improvement in our expected upper limit on $ZH$
%production.  
Our ANN configuration is 8 input variables, 17 hidden nodes, and
2 output nodes, such that the output distribution is two-dimensional (2-D),
with one axis separating $ZH$ and $Z+b\bar{b}$, and the other axis separating
$t\bar{t}$ and $ZH$.  The variables chosen, in order of importance for
minimizing the classification error, are the scalar sum of the transverse
energies of the jets and leptons composing the Higgs and $Z$ candidates
($H_T$), $\MET$, dijet mass, $\Delta R$ between first jet and $Z$ candidate,
$\Delta R$ between subleading jet and $Z$ candidate, $\Delta R$ between
leading and subleading jets, sphericity, which is a measure of how isotropic
the leptons and jets are, and $\eta$ of the subleading jet.  The most important
distributions are shown in Fig.~\ref{fig:dijet}.  
%We find
%that the data are modeled well in independent kinematic regions with an order
%of magnitude lower signal to background expected than our signal region.

%We optimize the ANN for
%$m_H =$ 120 GeV/$c^2$. We cross-check that the ANN input kinematics are
%modeled well in the data before the $b$-tag requirement and also plot the ANN
%outputs and expected and observed correlations to verify that the ANN
%properly recognizes the correlations between the input variables.  
%as is shown in Fig.~\ref{fig:pretag}.

%\begin{figure}[htbp]
%\includegraphics[width=0.45\textwidth]{pix/2Dshapes.eps}
%\caption{\label{fig:twoD}
%Shown is the 2-D ANN separation of $t\bar{t}$, $ZH$, and $Z+b\bar{b}$,
%normalized to unity for shape comparison. Inset on left is the one-tagged
%data, and on the right is two-tagged data.}
%\end{figure}

%\begin{figure}[htbp]
%\includegraphics[width=0.45\textwidth]{pix/mpdfMjj.eps}
%\caption{\label{fig:dijet}
%Shown is the dijet mass for the background, compared with data. The yellow
%and red lines are the expected $ZH$ signal for a rate of 50 times that of the
%standard model before and after the jet energies are corrected for the
%measured $\met$ of the event. }
%\end{figure}

%The observed distributions of ANN input
%kinematics and signal discriminant agree well with the predictions for the
%full event selection criteria. 
\begin{figure}[htbp]
\includegraphics[width=0.45\textwidth]{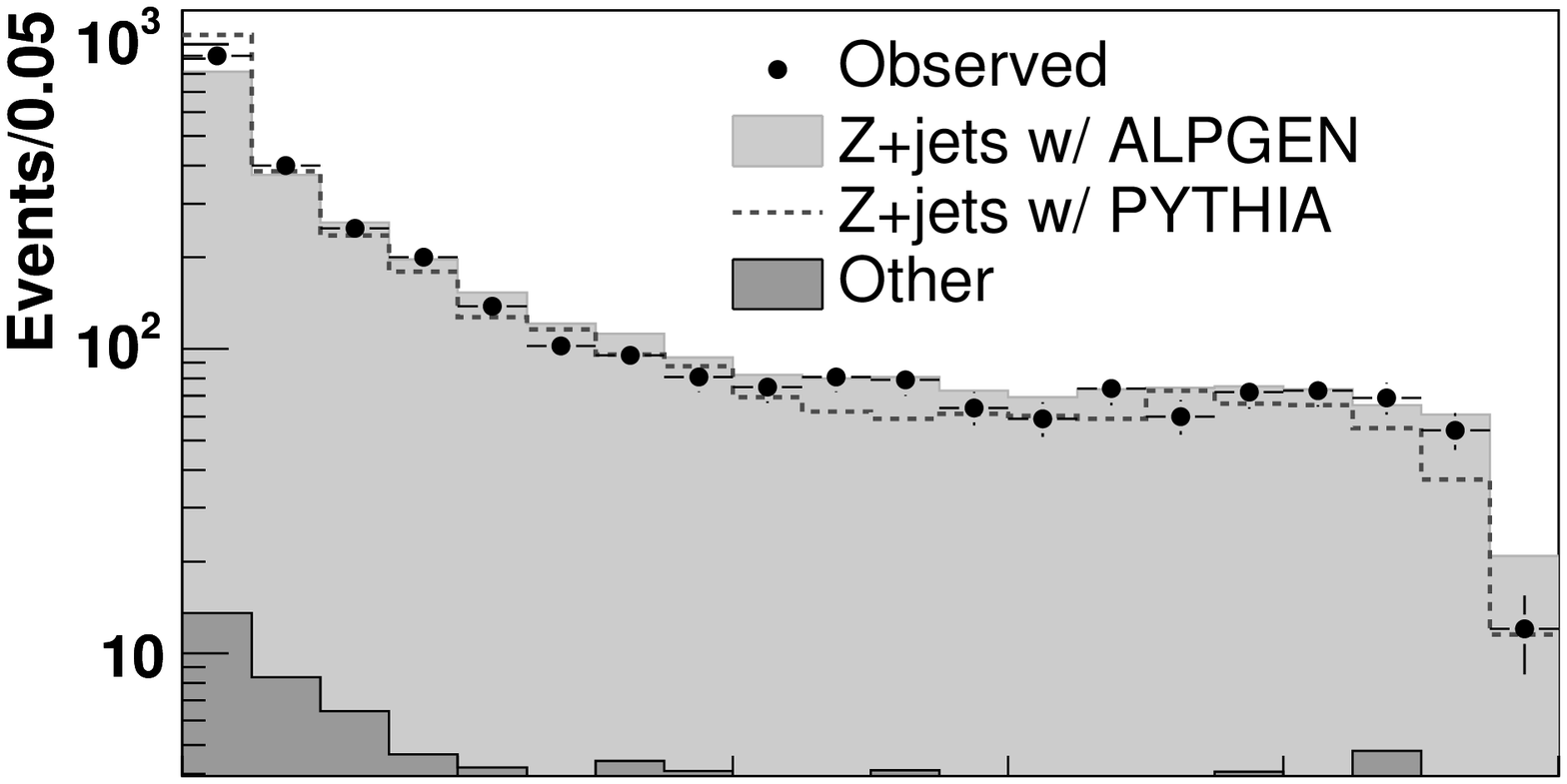}\\
\includegraphics[width=0.45\textwidth]{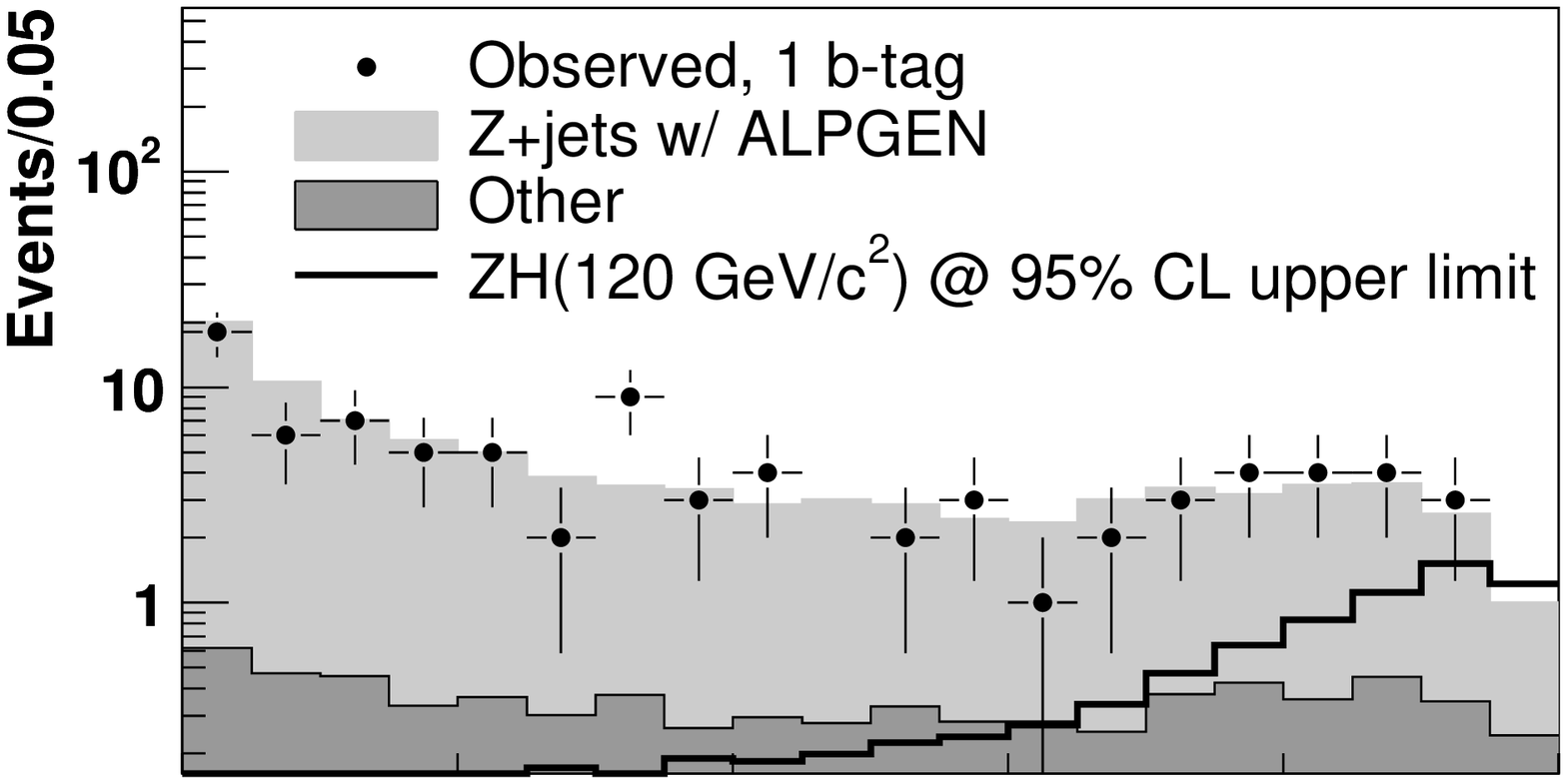}\\
\includegraphics[width=0.45\textwidth]{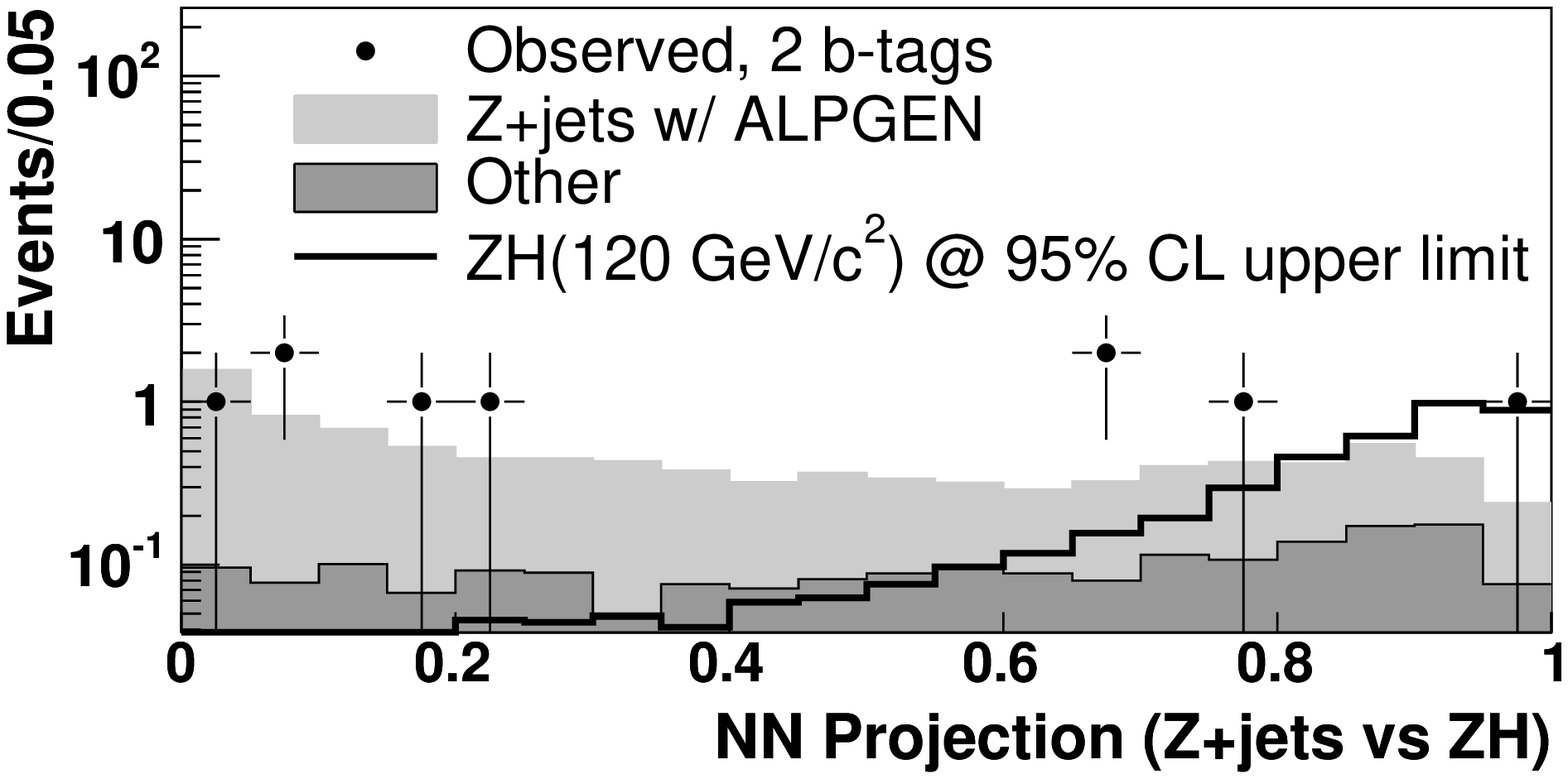}
\caption{\label{fig:tag_tt}
Expected and observed distributions for the ANN discriminant projected onto
the axis which discriminates $Z$+jets from $ZH$, after enriching signal by
selecting the most signal-like 25\% of events as determined by the $t\bar{t}$
vs. $ZH$ ANN output. The top plot is the ANN discriminant before $b$-tagging
requirements, demonstrating two models for the $Z$+jets background normalized
to data for shape comparison.  The middle and lower plots are for
single-tagged and double-tagged events, and are shown with a $ZH$ signal at
the level of the observed 95\% C.L. upper limit for $m_H =$ 120 GeV/$c^2$. }
\end{figure}

%We construct a likelihood for the expected background and signal in each bin
%of our 2-D ANN discriminant, and integrate over all parameters which affect
%these expectations in order to determine a posterior density function such
%that 95\% of the area represents the upper limit on the level of our
%$ZH$ signal. 
%Systematic
%uncertainties in the signal and relative background modeling are included as
%Gaussian constraints with correlations maintained as background components
%which are fluctuated in the signal fitting procedure.

The uncertainty on the amount of $Z+b\bar{b}$ and $Z+c\bar{c}$
background is taken to be 40\%~\cite{Abulencia:2006ce} and for the
$t\bar{t}$, $WZ$, and $ZZ$~\cite{Campbell:1999ah,Kidonakis:2003qe}, it is
20\%, including the uncertainties on the cross section and on the selection
efficiencies of these processes and the top quark mass uncertainty. The
uncertainty on the non-$Z$ background is 50\%. The uncertainty on the shape
of the background is evaluated by comparing $Z+b\bar{b}$ events between {\sc
pythia} and {\sc alpgen}. The signal shape uncertainty is evaluated by
varying the amount of initial and final state QCD
radiation~\cite{Abulencia:2005aj}, and by changing the parton distribution
functions using the 40 eigenvectors from {\sc CTEQ6}~\cite{Lai:1999wy}. We
evalute both rate and shape uncertainties for the signal and backgrounds by
varying the jet energy scale within its uncertainties
~\cite{Bhatti:2005ai}. In addition both signal and background estimates are
affected by the trigger efficiency uncertainty (1\%), and the luminosity
measurement uncertainty (6\%)~\cite{Acosta:2002hx}.  The $b$-tagging
efficiency has an uncertainty of 8\% for $b$-quark jets, 16\% for $c$-quark
jets and 13\% for l.p. jets. In double-tagged events, the
uncertainties are 16\%, 32\% and 24\%, respectively.  These values are
updates obtained from the procedure found in Ref.~\cite{Acosta:2004hw}.

The projections of the 2-D ANN signal discriminant are shown in
Fig.~\ref{fig:tag_tt}.  We analyze the binned 2-D ANN discriminant
distribution to test for a $ZH$ signal in the presence of SM backgrounds
using a Bayesian technique~\cite{Heinrich:2004tj} and marginalize over
variations in the systematic uncertainties.  The expected and observed upper
limits at the 95\% C.L. are shown in Table \ref{tab:limit}. Expected limits
are obtained by generating pseudo-experiments from the expected SM ANN shapes
to calculate the median $ZH$ contribution which could be excluded at the 95\%
level with no $ZH$ signal present.  Since backgrounds are smaller for
double-tagged events, we analyze them separately from single-tagged events,
resulting in a 20\% improvement in the expected limits.

The dominant systematic uncertainty is the $b$-quark identification
efficiency which accounts for 12\% of the total 14\% increase in the expected
limit due to systematic uncertainties.
\begin{table}
\caption{Upper limits at the 95\% C.L. on the cross-section of $\sigma(ZH)
\times {\cal B}(H\to b\bar{b})$.  Also shown is the expected and observed
ratio of the limit compared to the SM cross section. }
\label{tab:limit} 
\begin{center}
  \begin{tabular}{crrr}
   \hline
   \hline
% CDF 8784 :
   $m_H$     & Observed&  Observed              &    Expected              \\ 
   GeV/$c^2$ &  [pb]   & $\sigma$/$\sigma_{SM}$ &  $\sigma$/$\sigma_{SM}$  \\
   \hline                                      
   110       &  1.5    &      15                &    15                    \\
   115       &  1.4    &      17                &    16                    \\
   120       &  1.2    &      19                &    19                    \\
   130       &  1.2    &      30                &    28                    \\
   140       &  1.2    &      65                &    55                    \\
   150       &  1.2    &      160               &    140                   \\\hline
    \hline
   \end{tabular}
 \end{center}
\end{table}

In summary, we have extended the limits for a Higgs boson decaying to
$b\bar{b}$ produced in association with a $Z$ boson.  This is the first
Tevatron Run II search for a Higgs boson to use a multivariate approach to
separate signal and background kinematics, and results in a significant
improvement in Higgs sensitivity over previous analyses in this decay
mode~\cite{Abazov:2007pj}.  The observed event kinematics and ANN signal
discriminants show no significant excess above SM predictions. The
improvement in limits using our approach of two ANNs is a factor of 1.8
compared to a fit of the uncorrected dijet mass distribution alone.  This
result finds the best limit on Standard Model Higgs production in the most
favored Higgs boson mass range of less than 125 GeV/c$^2$ that was achieved with 1
fb$^{-1}$ of data as in Ref.~\cite{Efron:2007zz}.  Full sensitivity for
this mass range will be achieved by combining this analysis with other CDF
and D\O~Higgs search channels and the combined Tevatron dataset. 

%With a 1 fb$^{-1}$ dataset, we set 95\% C.L. upper limits on
%$\sigma(p\bar{p} \to ZH) \times {\cal B}(H\to b\bar{b})$ ranging from 1.5 pb
%at $m_H =$ 110 GeV$/c^2$ to 1.2 pb at $m_H =$ 150 GeV/$c^2$. 
  
%The statistics of this sample produce a limit
%which is 17 times larger than the SM cross section for $ZH$ production at
%$m_H =$ 115 GeV/$c^2$.

%This gap can be bridged by
%combining this channel with independent channels from CDF and D\O \ ,
%including future additional integrated luminosity, and by incorporating
%smarter techniques of lepton and $b$-jet identification and reconstruction.

We thank the Fermilab staff and the technical staffs of the participating 
institutions for their vital contributions. This work was supported by 
the U.S. Department of Energy and National Science Foundation; the 
Italian Istituto Nazionale di Fisica Nucleare; the Ministry of Education, 
Culture, Sports, Science and Technology of Japan; the Natural Sciences 
and Engineering Research Council of Canada; the National Science Council 
of the Republic of China; the Swiss National Science Foundation; the A.P. 
Sloan Foundation; the Bundesministerium f\"ur Bildung und Forschung, 
Germany; the Korean Science and Engineering Foundation and the Korean 
Research Foundation; the Science and Technology Facilities Council and the Royal 
Society, UK; the Institut National de Physique Nucleaire et 
Physique des Particules/CNRS; the Russian Foundation for Basic Research; 
the Ministerio de Educaci\'{o}n y Ciencia and Programa Consolider-Ingenio 2010, Spain;
the Slovak R\&D Agency; and the Academy of Finland.

%\begin{figure*}
%\includegraphics[width=0.45\textwidth]{pix/combinedElectronMuonData.eps}
%\caption{\label{fig:2ddata}
%Two-dimensional Neural Network discriminant of data.}
%\end{figure*}

%\begin{figure*}
%\includegraphics[width=0.45\textwidth]{pix/TOTALInvMass2Jet.eps}
%\caption{\label{fig:mjj}
%Dijet mass distribution of data with 1 $b$-tag requirement.}
%\end{figure*}

%This is the core of the PRL.  You can reference to this article \cite{Carena:1999xa} and to that thesis \cite{Eusebi_thesis_cite}. The only thing that does not get numbered is the abstract. Figure~\ref{myresult_fig} shows nothing.

% To include a figure named 2D_darker.eps just do the things below

% you should create/edit the file test.bib
\bibliography{zhllbb_prl}

\begin{thebibliography}{33}
\expandafter\ifx\csname natexlab\endcsname\relax\def\natexlab#1{#1}\fi
\expandafter\ifx\csname bibnamefont\endcsname\relax
  \def\bibnamefont#1{#1}\fi
\expandafter\ifx\csname bibfnamefont\endcsname\relax
  \def\bibfnamefont#1{#1}\fi
\expandafter\ifx\csname citenamefont\endcsname\relax
  \def\citenamefont#1{#1}\fi
\expandafter\ifx\csname url\endcsname\relax
  \def\url#1{\texttt{#1}}\fi
\expandafter\ifx\csname urlprefix\endcsname\relax\def\urlprefix{URL }\fi
\providecommand{\bibinfo}[2]{#2}
\providecommand{\eprint}[2][]{\url{#2}}

\bibitem[{\citenamefont{Higgs}(1964)}]{Higgs:1964ia}
\bibinfo{author}{\bibfnamefont{P.~W.} \bibnamefont{Higgs}},
  \bibinfo{journal}{Phys. Lett.} \textbf{\bibinfo{volume}{12}},
  \bibinfo{pages}{132} (\bibinfo{year}{1964}).

\bibitem[{\citenamefont{Guralnik et~al.}(1964)\citenamefont{Guralnik, Hagen,
  and Kibble}}]{Guralnik:1964eu}
\bibinfo{author}{\bibfnamefont{G.~S.} \bibnamefont{Guralnik}},
  \bibinfo{author}{\bibfnamefont{C.~R.} \bibnamefont{Hagen}}, \bibnamefont{and}
  \bibinfo{author}{\bibfnamefont{T.~W.~B.} \bibnamefont{Kibble}},
  \bibinfo{journal}{Phys. Rev. Lett.} \textbf{\bibinfo{volume}{13}},
  \bibinfo{pages}{585} (\bibinfo{year}{1964}).

\bibitem[{\citenamefont{Barate et~al.}(2003)}]{Barate:2003sz}
\bibinfo{author}{\bibfnamefont{R.}~\bibnamefont{Barate}} \bibnamefont{et~al.}
  (\bibinfo{collaboration}{LEP Working Group for Higgs boson searches}),
  \bibinfo{journal}{Phys. Lett.} \textbf{\bibinfo{volume}{B565}},
  \bibinfo{pages}{61} (\bibinfo{year}{2003}), \eprint{arXiv:hep-ex/0306033}.

\bibitem[{\citenamefont{Alcaraz et~al.}(2007)}]{Alcaraz:2007ri}
\bibinfo{author}{\bibfnamefont{J.}~\bibnamefont{Alcaraz}} \bibnamefont{et~al.}
  (\bibinfo{collaboration}{LEP Electroweak Working Group})
  (\bibinfo{year}{2007}), \bibinfo{note}{{{\it Precision Electroweak
  Measurements and Constraints on the Standard Model}}},
  \eprint{arXiv:hep-ex/07120929}.

\bibitem[{\citenamefont{Roszkowski et~al.}(2007)\citenamefont{Roszkowski,
  de~Austri, and Trotta}}]{Roszkowski:2006mi}
\bibinfo{author}{\bibfnamefont{L.}~\bibnamefont{Roszkowski}},
  \bibinfo{author}{\bibfnamefont{R.~R.} \bibnamefont{de~Austri}},
  \bibnamefont{and} \bibinfo{author}{\bibfnamefont{R.}~\bibnamefont{Trotta}},
  \bibinfo{journal}{J. High Energy Phys.} \textbf{\bibinfo{volume}{04}},
  \bibinfo{pages}{084} (\bibinfo{year}{2007}), \eprint{arXiv:hep-ph/0611173}.

\bibitem[{\citenamefont{Abazov et~al.}(2007)}]{Abazov:2007pj}
\bibinfo{author}{\bibfnamefont{V.~M.} \bibnamefont{Abazov}}
  \bibnamefont{et~al.} (\bibinfo{collaboration}{D\O \ Collaboration}),
  \bibinfo{journal}{Phys. Lett.} \textbf{\bibinfo{volume}{B655}},
  \bibinfo{pages}{209} (\bibinfo{year}{2007}), \eprint{arXiv:hep-ex/07042000}.

\bibitem[{\citenamefont{Aaltonen et~al.}(2008{\natexlab{a}})}]{Aaltonen:2007wx}
\bibinfo{author}{\bibfnamefont{T.}~\bibnamefont{Aaltonen}} \bibnamefont{et~al.}
  (\bibinfo{collaboration}{CDF}), \bibinfo{journal}{Phys. Rev. Lett.}
  \textbf{\bibinfo{volume}{100}}, \bibinfo{pages}{041801}
  (\bibinfo{year}{2008}{\natexlab{a}}), \eprint{arXiv:hep-ex/07104363}.

\bibitem[{\citenamefont{Aaltonen et~al.}(2008{\natexlab{b}})}]{Aaltonen:2008mi}
\bibinfo{author}{\bibfnamefont{T.}~\bibnamefont{Aaltonen}} \bibnamefont{et~al.}
  (\bibinfo{collaboration}{CDF}), \bibinfo{journal}{Phys. Rev. Lett.}
  \textbf{\bibinfo{volume}{100}}, \bibinfo{pages}{211801}
  (\bibinfo{year}{2008}{\natexlab{b}}), \eprint{arXiv:hep-ex/08020432}.

\bibitem[{\citenamefont{Abulencia
  et~al.}(2006{\natexlab{a}})}]{Abulencia:2006aj}
\bibinfo{author}{\bibfnamefont{A.}~\bibnamefont{Abulencia}}
  \bibnamefont{et~al.} (\bibinfo{collaboration}{CDF Collaboration}),
  \bibinfo{journal}{Phys. Rev. Lett.} \textbf{\bibinfo{volume}{97}},
  \bibinfo{pages}{081802} (\bibinfo{year}{2006}{\natexlab{a}}),
  \eprint{arXiv:hep-ex/0605124}.

\bibitem[{\citenamefont{Abazov et~al.}(2006{\natexlab{a}})}]{Abazov:2005un}
\bibinfo{author}{\bibfnamefont{V.~M.} \bibnamefont{Abazov}}
  \bibnamefont{et~al.} (\bibinfo{collaboration}{D\O \ Collaboration}),
  \bibinfo{journal}{Phys. Rev. Lett.} \textbf{\bibinfo{volume}{96}},
  \bibinfo{pages}{011801} (\bibinfo{year}{2006}{\natexlab{a}}),
  \eprint{arXiv:hep-ex/0508054}.

\bibitem[{\citenamefont{Abazov et~al.}(2005)}]{Abazov:2004jy}
\bibinfo{author}{\bibfnamefont{V.~M.} \bibnamefont{Abazov}}
  \bibnamefont{et~al.} (\bibinfo{collaboration}{D\O \ Collaboration}),
  \bibinfo{journal}{Phys. Rev. Lett.} \textbf{\bibinfo{volume}{94}},
  \bibinfo{pages}{091802} (\bibinfo{year}{2005}),
  \eprint{arXiv:hep-ex/0410062}.

\bibitem[{\citenamefont{Abazov et~al.}(2006{\natexlab{b}})}]{Abazov:2006pt}
\bibinfo{author}{\bibfnamefont{V.~M.} \bibnamefont{Abazov}}
  \bibnamefont{et~al.} (\bibinfo{collaboration}{D\O \ Collaboration}),
  \bibinfo{journal}{Phys. Rev. Lett.} \textbf{\bibinfo{volume}{97}},
  \bibinfo{pages}{161803} (\bibinfo{year}{2006}{\natexlab{b}}),
  \eprint{arXiv:hep-ex/0607022}.

\bibitem[{\citenamefont{Abazov et~al.}(2006{\natexlab{c}})}]{Abazov:2006hn}
\bibinfo{author}{\bibfnamefont{V.~M.} \bibnamefont{Abazov}}
  \bibnamefont{et~al.} (\bibinfo{collaboration}{D\O \ Collaboration}),
  \bibinfo{journal}{Phys. Rev. Lett.} \textbf{\bibinfo{volume}{97}},
  \bibinfo{pages}{151804} (\bibinfo{year}{2006}{\natexlab{c}}),
  \eprint{arXiv:hep-ex/0607032}.

\bibitem[{\citenamefont{Acosta et~al.}(2005{\natexlab{a}})}]{Acosta:2004yw}
\bibinfo{author}{\bibfnamefont{D.}~\bibnamefont{Acosta}} \bibnamefont{et~al.}
  (\bibinfo{collaboration}{CDF Collaboration}), \bibinfo{journal}{Phys. Rev. D}
  \textbf{\bibinfo{volume}{71}}, \bibinfo{pages}{032001}
  (\bibinfo{year}{2005}{\natexlab{a}}), \eprint{arXiv:hep-ex/0412071}.

\bibitem[{coo()}]{coord}
\bibinfo{note}{In the CDF coordinate system, the z-axis is the proton
  direction, $\theta$ is the polar angle, and $\phi$ is the azimuthal angle.
  Pseudorapidity $\eta$ is defined as --ln[tan ($\theta$/2)]. The transverse
  momentum of a particle is $p_T =$ $p$ sin $\theta$, where $p$ is its
  momentum. The transverse energy $E_T$ is an analogous quantity defined from
  calorimeter energies and coordinates as $E_T = E$ sin $\theta$. The missing
  transverse energy $\MET$ is defined as - $|\sum{ E^i_T \hat{n}_{i} }|$ where
  $\hat{n}_{i}$ is the radial unit vector pointing from the interaction point
  to the energy deposition in a given calorimeter cell, $i$. $\Delta$R is
  defined as $\sqrt{ {(\Delta\phi)}^2 + {(\Delta\eta)}^2 }$ and used to specify
  particle separation.}

\bibitem[{\citenamefont{Catani et~al.}(2003)\citenamefont{Catani, de~Florian,
  Grazzini, and Nason}}]{Catani:2003zt}
\bibinfo{author}{\bibfnamefont{S.}~\bibnamefont{Catani}},
  \bibinfo{author}{\bibfnamefont{D.}~\bibnamefont{de~Florian}},
  \bibinfo{author}{\bibfnamefont{M.}~\bibnamefont{Grazzini}}, \bibnamefont{and}
  \bibinfo{author}{\bibfnamefont{P.}~\bibnamefont{Nason}}, \bibinfo{journal}{J.
  High Energy Phys.} \textbf{\bibinfo{volume}{0307}}, \bibinfo{pages}{028}
  (\bibinfo{year}{2003}), \eprint{arXiv:hep-ph/0306211}.

\bibitem[{\citenamefont{Djouadi et~al.}(1998)\citenamefont{Djouadi, Kalinowski,
  and Spira}}]{Djouadi:1997yw}
\bibinfo{author}{\bibfnamefont{A.}~\bibnamefont{Djouadi}},
  \bibinfo{author}{\bibfnamefont{J.}~\bibnamefont{Kalinowski}},
  \bibnamefont{and} \bibinfo{author}{\bibfnamefont{M.}~\bibnamefont{Spira}},
  \bibinfo{journal}{Comput. Phys. Commun.} \textbf{\bibinfo{volume}{108}},
  \bibinfo{pages}{56} (\bibinfo{year}{1998}), \eprint{arXiv:hep-ph/9704448}.

\bibitem[{\citenamefont{Efron}(2007)}]{Efron:2007zz}
\bibinfo{author}{\bibfnamefont{J.~Z.} \bibnamefont{Efron}}, Ph.D. thesis,
  \bibinfo{school}{The Ohio State University} (\bibinfo{year}{2007}),
  \bibinfo{note}{[Fermilab Report No. FERMILAB-THESIS-2007-20, 2007], Fig. 8.1
  shows expected (dashed) and observed (solid) limits for other CDF analysis
  channels}.

\bibitem[{\citenamefont{Bhatti et~al.}(2006)}]{Bhatti:2005ai}
\bibinfo{author}{\bibfnamefont{A.}~\bibnamefont{Bhatti}} \bibnamefont{et~al.},
  \bibinfo{journal}{Nucl. Instrum. Methods} \textbf{\bibinfo{volume}{A566}},
  \bibinfo{pages}{375} (\bibinfo{year}{2006}), \eprint{arXiv:hep-ex/0510047}.

\bibitem[{\citenamefont{Acosta et~al.}(2005{\natexlab{b}})}]{Acosta:2004hw}
\bibinfo{author}{\bibfnamefont{D.}~\bibnamefont{Acosta}} \bibnamefont{et~al.}
  (\bibinfo{collaboration}{CDF Collaboration}), \bibinfo{journal}{Phys. Rev. D}
  \textbf{\bibinfo{volume}{71}}, \bibinfo{pages}{052003}
  (\bibinfo{year}{2005}{\natexlab{b}}), \eprint{arXiv:hep-ex/0410041}.

\bibitem[{\citenamefont{Neu}(2006)}]{Neu:2006rs}
\bibinfo{author}{\bibfnamefont{C.}~\bibnamefont{Neu}}
  (\bibinfo{collaboration}{CDF Collaboration}) (\bibinfo{year}{2006}),
  \bibinfo{note}{in Proceedings of the International Workshop on Top Quark
  Physics, January 12-15, 2006, Coimbra, Portugal, Proceedings of Science,
  Trieste, Italy, 2006.}

\bibitem[{\citenamefont{Mangano et~al.}(2003)\citenamefont{Mangano, Moretti,
  Piccinini, Pittau, and Polosa}}]{Mangano:2002ea}
\bibinfo{author}{\bibfnamefont{M.~L.} \bibnamefont{Mangano}},
  \bibinfo{author}{\bibfnamefont{M.}~\bibnamefont{Moretti}},
  \bibinfo{author}{\bibfnamefont{F.}~\bibnamefont{Piccinini}},
  \bibinfo{author}{\bibfnamefont{R.}~\bibnamefont{Pittau}}, \bibnamefont{and}
  \bibinfo{author}{\bibfnamefont{A.~D.} \bibnamefont{Polosa}},
  \bibinfo{journal}{J. High Energy Phys.} \textbf{\bibinfo{volume}{07}},
  \bibinfo{pages}{001} (\bibinfo{year}{2003}), \eprint{arXiv:hep-ph/0206293}.

\bibitem[{\citenamefont{Corcella et~al.}(2002)}]{Corcella:2002jc}
\bibinfo{author}{\bibfnamefont{G.}~\bibnamefont{Corcella}} \bibnamefont{et~al.}
  (\bibinfo{year}{2002}), \bibinfo{note}{{\sc herwig} 6.5 release note},
  \eprint{arXiv:hep-ph/0210213}.

\bibitem[{\citenamefont{Sjostrand et~al.}(2001)\citenamefont{Sjostrand,
  Lonnblad, and Mrenna}}]{Sjostrand:2001yu}
\bibinfo{author}{\bibfnamefont{T.}~\bibnamefont{Sjostrand}},
  \bibinfo{author}{\bibfnamefont{L.}~\bibnamefont{Lonnblad}}, \bibnamefont{and}
  \bibinfo{author}{\bibfnamefont{S.}~\bibnamefont{Mrenna}}
  (\bibinfo{year}{2001}), \bibinfo{note}{{\sc pythia} 6.2: Physics and manual},
  \eprint{arXiv:hep-ph/0108264}.

\bibitem[{\citenamefont{Abulencia
  et~al.}(2006{\natexlab{b}})}]{Abulencia:2006ce}
\bibinfo{author}{\bibfnamefont{A.}~\bibnamefont{Abulencia}}
  \bibnamefont{et~al.} (\bibinfo{collaboration}{CDF Collaboration}),
  \bibinfo{journal}{Phys. Rev. D} \textbf{\bibinfo{volume}{74}},
  \bibinfo{pages}{032008} (\bibinfo{year}{2006}{\natexlab{b}}),
  \eprint{arXiv:hep-ex/0605099}.

\bibitem[{\citenamefont{Hornik et~al.}(1989)\citenamefont{Hornik, Stinchcombe,
  and White}}]{DBLP:journals/nn/HornikSW89}
\bibinfo{author}{\bibfnamefont{K.}~\bibnamefont{Hornik}},
  \bibinfo{author}{\bibfnamefont{M.~B.} \bibnamefont{Stinchcombe}},
  \bibnamefont{and} \bibinfo{author}{\bibfnamefont{H.}~\bibnamefont{White}},
  \bibinfo{journal}{Neural Networks} \textbf{\bibinfo{volume}{2}},
  \bibinfo{pages}{359} (\bibinfo{year}{1989}), \bibinfo{note}{we use the MLPfit
  program which can be found online},
  \urlprefix\url{http://schwind.web.cern.ch/schwind/MLPfit.html}.

\bibitem[{\citenamefont{Peterson et~al.}(1994)\citenamefont{Peterson,
  Rognvaldsson, and Lonnblad}}]{Peterson:1993nk}
\bibinfo{author}{\bibfnamefont{C.}~\bibnamefont{Peterson}},
  \bibinfo{author}{\bibfnamefont{T.}~\bibnamefont{Rognvaldsson}},
  \bibnamefont{and} \bibinfo{author}{\bibfnamefont{L.}~\bibnamefont{Lonnblad}},
  \bibinfo{journal}{Comput. Phys. Commun.} \textbf{\bibinfo{volume}{81}},
  \bibinfo{pages}{185} (\bibinfo{year}{1994}), \bibinfo{note}{we use a {\sc
  root}-based version of {\sc jetnet} called Root\_Jetnet which can be found
  online},
  \urlprefix\url{http://www.hep.uiuc.edu/home/catutza/root_to_jetnet/rtj.html}.

\bibitem[{\citenamefont{Campbell and Ellis}(1999)}]{Campbell:1999ah}
\bibinfo{author}{\bibfnamefont{J.~M.} \bibnamefont{Campbell}} \bibnamefont{and}
  \bibinfo{author}{\bibfnamefont{R.~K.} \bibnamefont{Ellis}},
  \bibinfo{journal}{Phys. Rev. D} \textbf{\bibinfo{volume}{60}},
  \bibinfo{pages}{113006} (\bibinfo{year}{1999}),
  \eprint{arXiv:hep-ph/9905386}.

\bibitem[{\citenamefont{Kidonakis and Vogt}(2003)}]{Kidonakis:2003qe}
\bibinfo{author}{\bibfnamefont{N.}~\bibnamefont{Kidonakis}} \bibnamefont{and}
  \bibinfo{author}{\bibfnamefont{R.}~\bibnamefont{Vogt}},
  \bibinfo{journal}{Phys. Rev. D} \textbf{\bibinfo{volume}{68}},
  \bibinfo{pages}{114014} (\bibinfo{year}{2003}),
  \eprint{arXiv:hep-ph/0308222}.

\bibitem[{\citenamefont{Abulencia
  et~al.}(2006{\natexlab{c}})}]{Abulencia:2005aj}
\bibinfo{author}{\bibfnamefont{A.}~\bibnamefont{Abulencia}}
  \bibnamefont{et~al.} (\bibinfo{collaboration}{CDF Collaboration}),
  \bibinfo{journal}{Phys. Rev. D} \textbf{\bibinfo{volume}{73}},
  \bibinfo{pages}{032003} (\bibinfo{year}{2006}{\natexlab{c}}),
  \eprint{arXiv:hep-ex/0510048}.

\bibitem[{\citenamefont{Lai et~al.}(2000)}]{Lai:1999wy}
\bibinfo{author}{\bibfnamefont{H.~L.} \bibnamefont{Lai}} \bibnamefont{et~al.}
  (\bibinfo{collaboration}{CTEQ}), \bibinfo{journal}{Eur. Phys. J. C}
  \textbf{\bibinfo{volume}{12}}, \bibinfo{pages}{375} (\bibinfo{year}{2000}),
  \eprint{arXiv:hep-ph/9903282}.

\bibitem[{\citenamefont{Acosta et~al.}(2002)}]{Acosta:2002hx}
\bibinfo{author}{\bibfnamefont{D.}~\bibnamefont{Acosta}} \bibnamefont{et~al.},
  \bibinfo{journal}{Nucl. Instrum. Methods} \textbf{\bibinfo{volume}{A494}},
  \bibinfo{pages}{57} (\bibinfo{year}{2002}).

\bibitem[{\citenamefont{Heinrich et~al.}(2004)\citenamefont{Heinrich, Blocker,
  Conway, Demortier, Lyons, Punzi, and Sinervo}}]{Heinrich:2004tj}
\bibinfo{author}{\bibfnamefont{J.}~\bibnamefont{Heinrich}},
  \bibinfo{author}{\bibfnamefont{C.}~\bibnamefont{Blocker}},
  \bibinfo{author}{\bibfnamefont{J.}~\bibnamefont{Conway}},
  \bibinfo{author}{\bibfnamefont{L.}~\bibnamefont{Demortier}},
  \bibinfo{author}{\bibfnamefont{L.}~\bibnamefont{Lyons}},
  \bibinfo{author}{\bibfnamefont{G.}~\bibnamefont{Punzi}}, \bibnamefont{and}
  \bibinfo{author}{\bibfnamefont{P.}~\bibnamefont{Sinervo}}
  (\bibinfo{year}{2004}), \bibinfo{note}{{Interval estimation in the presence
  of nuisance parameters. 1. Bayesian approach}},
  \eprint{arXiv:physics/0409129v1}.

\end{thebibliography}

\end{document}